\documentclass[12pt]{article}
\usepackage{graphics}
\begin{document}
%
\begin{center}
{\bf \LARGE Three-Pion Interferometry Results from Central Pb+Pb Collisions at
158 $A$ GeV/c}
\end{center}
\vspace*{0.7cm}
{\small \noindent
M.M.~Aggarwal,$^{1}$ 
A.~Agnihotri,$^{2}$ 
Z.~Ahammed,$^{3}$ 
A.L.S.~Angelis,$^{4}$  
V.~Antonenko,$^{5}$  
V.~Arefiev,$^{6}$ 
V.~Astakhov,$^{6}$ 
V.~Avdeitchikov,$^{6}$ 
T.C.~Awes,$^{7}$ 
P.V.K.S.~Baba,$^{8}$
% ...
}
\\ \vspace*{-0.1cm}
\begin{center} (WA98 collaboration) \end{center}
\vspace*{0.2cm}
%
{\small
{$^{1}$~University of Panjab, Chandigarh 160014, India} \\
{$^{2}$~University of Rajasthan, Jaipur 302004, Rajasthan, India} \\
{$^{3}$~Variable Energy Cyclotron Centre,  Calcutta 700 064, India} \\
{$^{4}$~University of Geneva, CH-1211 Geneva 4,Switzerland} \\
{$^{5}$~RRC Kurchatov Institute, RU-123182 Moscow, Russia} \\
{$^{6}$~Joint Institute for Nuclear Research, RU-141980 Dubna, Russia} \\
{$^{7}$~Oak Ridge National Laboratory, Oak Ridge, Tennessee 37831-6372, USA} \\
{$^{8}$~University of Jammu, Jammu 180001, India} \\
{$^{9}$~University of M{\"u}nster, D-48149 M{\"u}nster, Germany} \\
{$^{10}$~SUBATECH, Ecole des Mines, Nantes, France} \\
{$^{11}$~Gesellschaft f{\"u}r Schwerionenforschung (GSI), D-64220 Darmstadt, Germany} \\
{$^{12}$~Universiteit Utrecht/NIKHEF, NL-3508 TA Utrecht, The Netherlands} \\
{$^{13}$~Lund University, SE-221 00 Lund, Sweden} \\
{$^{14}$~University of Tsukuba, Ibaraki 305, Japan} \\
{$^{15}$~Nuclear Physics Institute, CZ-250 68 Rez, Czech Rep.} \\
{$^{16}$~KVI, University of Groningen, NL-9747 AA Groningen, The Netherlands} \\
{$^{17}$~Institute for Nuclear Studies, 00-681 Warsaw, Poland} \\
{$^{18}$~MIT, Cambridge, MA 02139, USA} \\
{$^{19}$~Institute of Physics, 751-005  Bhubaneswar, India} \\
{$^{20}$~University of Tennessee, Knoxville, Tennessee 37966, USA} \\
}
%
\normalsize
%
\abstract{
Three-particle correlations have been measured for identified $\pi^-$ from
central 158 A GeV Pb+Pb collisions  by the WA98 experiment at CERN.
A substantial contribution of the genuine three-body correlation has been
found as expected for a mainly chaotic and symmetric source.
}
%
%
\\ \vspace{1.0cm} \\
In nuclear and particle physics, the study of Bose-Einstein correlations
between identical particles is widely used.
It is an essential tool to obtain information on the size and time
evolution of expanding systems 
created in heavy ion collisions \cite{ZAK}.
In particular, two-particle  Bose-Einstein interferometry has
been used for a sophisticated
analysis
of the dynamical evolution of the freeze out volume via selection on the
transverse momentum and rapidity of the correlated particles.

Three-particle interferometry measurements can provide
additional information
on the space-time emission which is not accessible by
two-particle interferometry\cite{ANDR,LORS,HEINZ,HEISEL}. 
In particular, if the emission is fully chaotic, the three-particle
interference study gives access to the phase of the source function's
Fourier transform, which is affected by the emission asymmetry.
These asymmetries may be induced by source geometry, flow or resonance
decays.
If the source is not completely chaotic, as is likely to be the case,
the interpretation is more difficult.
Nevertheless, a comparison
of the three-particle to the two-particle result 
allows to extract the relative strength of the true three-body correlation
and can in principle remove
obscuring effects, such as background from misidentified tracks
or long-lived decay products, and therefore provide information
more directly 
about the degree of coherence of the emission source\cite{HEINZ}.
In this letter we
present first results from three-particle interferometry in central
$^{208}$Pb+$^{208}$Pb collisions at the CERN SPS.

The fixed target experiment WA98 \cite{PROP} combined large acceptance
photon detectors with a two arm charged particle tracking spectrometer.
The incident 158 $A$ GeV Pb beam impinged on a Pb target near the entrance
of a large dipole magnet.
The results presented here have been obtained from an analysis of the
1995 data set.
These data were taken with the most central triggers corresponding to
about 10\% of the minimum bias cross section of 6190 mb.
%
The $\pi^-$ measurements were obtained with data
from the negative particle tracking arm of the spectrometer.
This tracking arm consisted of six multistep avalanche chambers with
optical readout \cite{LUX}.
A time of flight detector with a time resolution better than 120 ps
allowed for particle identification.
The rapidity acceptance ranged
from $y=2.1$ to 3.1 with an average at 2.7, close to mid-rapidity 
which was 2.9.
The momentum resolution of the spectrometer was $\Delta p/p=0.005$ at
$p=1.5$ GeV/c.
Severe track quality cuts were applied, resulting in a final sample of
4.2$\times10^6$ $\pi^-$, providing 7.2$\times10^6$ pairs and
8.2$\times10^6$ triplets.
The $\pi^-\pi^-$ correlation analysis has been reported elsewhere
\cite{BE}.

The most common use of hadron interferometry concerns the study of pairs
of identical particles.
The two-particle correlation function can be defined as
\begin{displaymath}
C_2({\mathbf{p}}_{1},{\mathbf{p}}_{2}) =
\frac{ d^{2}N({\mathbf{p}}_{1},{\mathbf{p}}_{2}) /
d{\mathbf{p}}_{1}d{\mathbf{p}}_{2} }
                { dN({\mathbf{p}}_{1}) / d{\mathbf{p}}_{1} \cdot
dN({\mathbf{p}}_{2})/d{\mathbf{p}}_{2}}
\end{displaymath}
where ${\mathbf{p}}_{1}$ and ${\mathbf{p}}_{2}$ are the 3--momenta of the
correlated particles.
The product of single particle distributions in the denominator is
usually obtained
experimentally by a mixed event technique whereas the pair distribution in
the numerator is constructed from 
all pair combinations of
identical particles found in each event.
$C_2$ is normalized to unity far away from the interference region.
In the plane wave approximation for chaotic sources of identical particles
$C_2$ can be written as \cite{CHAP} 
\begin{equation}
C_2({\mathbf{p}}_{1},{\mathbf{p}}_{2})=1\pm|F_{12}|^2
\label{eq:c2}
\end{equation}
with the + (-) sign for bosons (fermions).
$F_{12}$ is the Fourier transform of the 
space-time source function $S(x,k_{12})$
\begin{equation}
|F_{12}|^2 = \frac{|\int\!\:d^4x\;S(x,k_{12})\;\exp[iq_{12}x]|^2}{|\int\!
\:d^4x\; S(x,k_{12})|^2}
\label{eq:fourier}
\end{equation}
with $q_{12}=p_{1}-p_{2}$, the 4--momentum difference of the two
particles, and $k_{12}=(p_{1}+p_{2})/2$.
$F_{12}$ is unity as $q_{12}\rightarrow0$.
Thus the production of identical bosons is  enhanced for pairs 
created close together in phase space with a small momentum difference
$Q_{12}\equiv\sqrt{-q_{12}^2}$.

Typically $\pi\pi$ correlation data are fit with a
Gaussian form for $|F_{12}|^2$ 
\begin{equation}
C_2=1+\lambda\: \exp[-Q_{12}^2R^2]
\label{eq:c2gauss}
\end{equation}
or an exponential form
\begin{equation}
C_2=1+\lambda\: \exp[-2Q_{12}R]
\label{eq:c2exp}
\end{equation}
where the parameter $\lambda$ is inserted 
to take into account the possibility that the source may
not be fully chaotic and also that any wrongly reconstructed tracks, or
tracks coming from decays of long-lived resonances, will
dilute the correlations in the data.

The measurement of $C_2$ gives access to the radius $R$ of the
source, but not to the phase $\phi_{12}$ contained in $F_{12} \equiv
|F_{12}|\exp[i\phi_{12}]$ 
since $C_2$ is only a function of
the square of the Fourier transform of the source distribution $S$.
By contrast, the three-boson interference produced by a fully chaotic
source is sensitive to the phase information of the Fourier transform of
the source emission function.
Indeed, for a chaotic source the three-body 
correlation function $C_3$, which is
\begin{displaymath}
C_3({\mathbf{p}}_{1},{\mathbf{p}}_{2},{\mathbf{p}}_{3}) =
\frac{ d^{3}N({\mathbf{p}}_{1},{\mathbf{p}}_{2},{\mathbf{p}}_{3}) /
d{\mathbf{p}}_{1}d{\mathbf{p}}_{2}d{\mathbf{p}}_{3}}
{dN({\mathbf{p}}_{1})/d{\mathbf{p}}_{1} \cdot
dN({\mathbf{p}}_{2})/d{\mathbf{p}}_{2} \cdot
dN({\mathbf{p}}_{3})/d{\mathbf{p}}_{3} } 
\end{displaymath}
can be written \cite{ANDR,LORS,HEINZ,HEISEL} as
\begin{equation}
C_3=1+|F_{12}|^2+|F_{23}|^2+|F_{31}|^2+2\cdot \mbox{Re}\{ F_{12}\cdot 
F_{23}\cdot F_{31}\}
\label{eq:c3}
\end{equation}
where $F_{ij}$ is the Fourier transform for the pair $ij$ contained in the
triplet $123$. The genuine three-body correlation in $C_3$ is the term
$2\cdot \mbox{Re}\{ F_{12}\cdot F_{23}\cdot F_{31}\}$.

With $F_{ij}\equiv|F_{ij}|\exp[i\phi_{ij}]$ and
$W\equiv \cos(\phi_{12}+\phi_{23}+\phi_{31})$ one may rewrite
\begin{equation}
2\cdot \mbox{Re}\{ F_{12}\cdot F_{23}\cdot F_{31}\} =2\cdot |F_{12}|\cdot
|F_{23}|\cdot
|F_{31}|\cdot W.
\label{eq:cross}
\end{equation}
Having determined $|F_{ij}|$ from the pair correlation function $C_2$, the
measurement of $C_3$ gives direct access to $W$, the 
cosine of the sum of the three
phases of the Fourier transforms, and hence provides complementary
information on the shape of the source.
Indeed, $W$ is related to the odd space-time moments of the source which
generate the phases $\phi_{ij}$~\cite{HEINZ}.
For example, if the source is fully chaotic and symmetric, $F_{ij}$ are
real,
$\phi_{12}=\phi_{23}=\phi_{31}=0$, and $W$ is equal to 1.
$C_3$ is then fully determined by $C_2$, and the genuine three-particle
correlation is maximum.
%
If the source is not fully chaotic, more 
complicated expressions should be used, mixing the effects of
fully chaotic and coherent sources \cite{HEINZ}.
Nevertheless, Eq.~\ref{eq:cross} 
is valid in general with
$W$ interpreted as a factor expressing
the relative strength of the true three-body correlation compared
to that expected for a fully symmetric chaotic source.
Consequently, measuring $W$ values less than 1 does not allow to
differentiate between
asymmetries or coherence in the source.

Assuming a Gaussian form of the source function, which for two bosons
leads to a correlation function described by Eq.~\ref{eq:c2gauss}, one
obtains
\begin{equation}
C_3=1+\lambda\sum_{ij=12,23,31}\exp[-Q_{ij}^2R^2]
+2\lambda^{3/2}\exp[-Q_3^2R^2/2]\cdot W 
\label{eq:c3gauss}
\end{equation}
with $Q_3^2\equiv Q_{12}^2+Q_{23}^2+Q_{31}^2$.
If the exponential form of the Fourier transform is 
assumed instead (Eq.~\ref{eq:c2exp}), one obtains 
\begin{equation}
C_3=1+\lambda\sum_{ij=12,23,31}\exp[-2Q_{ij}R]
+2\lambda^{3/2}\exp[-(Q_{12}+Q_{23}+Q_{31})R]\cdot W 
\label{eq:c3exp}
\end{equation}
Inserting  
the values of $\lambda$ and $R$ obtained
from the two-particle analysis into Eqs.~\ref{eq:c3gauss} or
\ref{eq:c3exp}, one can extract the three-particle 
strength information $W$.

\begin{figure}[h]
\vspace{2.0cm}
\hspace{4.5cm}
\resizebox{0.5\textwidth}{!}{%
  \includegraphics{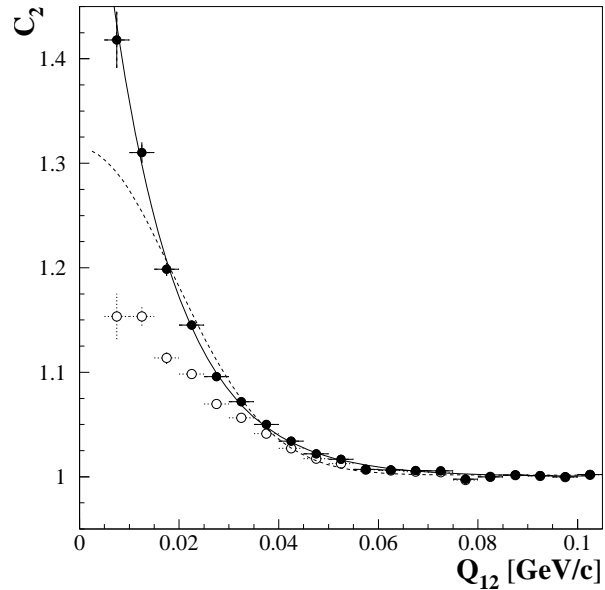}
}
\caption{The measured $\pi^-\pi^-$ correlation function $C_2$ (full symbols)
fit with an exponential (solid curve) or Gaussian form  (dashed curve) 
corrected for resolution. The empty symbols show the data before Coulomb and 
resolution corrections. Only
statistical errors are shown.}
\label{fig:figure1}
\end{figure}

The one-dimensional correlation function $C_2$, analyzed in terms of
$Q_{12}$, is shown in Fig.~\ref{fig:figure1}.
The two-track resolution of the spectrometer (2 cm) is dealt with by
application of a proximity cut to each track pair.
The data are corrected for the Coulomb and stong final state
interactions in an iterative way \cite{PRATT}, taking into account the
source size obtained in the fit.
The Gamow correction was abandoned as we found that it overcorrects the
data for
$Q_{12}$ in the range of 0.1 to 0.3 GeV/c.
The effect of the experimental resolution,
which is estimated by a full Monte-Carlo, can be approximated in the
interference region by a Gaussian of constant $\sigma$ of 7 MeV/c both
for $Q_{12}$ and $Q_3$. 
It is taken into account by replacing the 
function $C_2(Q_{12})$ used to fit the non-corrected data by
$$C_2^{rc}(Q_{12}) =
\int\! r(Q_{12},Q'_{12})\:C_2(Q'_{12})\:dQ'_{12},$$ 
which is the convolution of $C_2(Q_{12})$ with the resolution function
$r(Q_{12},Q'_{12})$ of the spectrometer.
For display purposes,
Fig.~\ref{fig:figure1} is obtained by
multiplying each data point by $C_2^{rc}(Q_{12})/C_2(Q_{12})$.
Fitting the corrected data with $C_2$ gives the same results
as fitting the non-corrected data with $C_2^{rc}$.
The correlation function $C_2$
is seen to be non-Gaussian. Instead,
it is better represented by an exponential form\cite{BE}. 
For the $C_3$ analysis, an accurate description of the shape of $C_2$
is essential since 
the estimate of the $W$ factor extracted from $C_3$ depends on it.
The exponential fit (Eq.~\ref{eq:c2exp}) 
yielding $R=7.29\pm0.11$ fm and
$\lambda=0.753\pm0.013$ is shown in Fig.~\ref{fig:figure1}.

\begin{figure}[t]
\vspace{-3.0cm}
\hspace{4.5cm}
\resizebox{0.5\textwidth}{!}{%
  \includegraphics{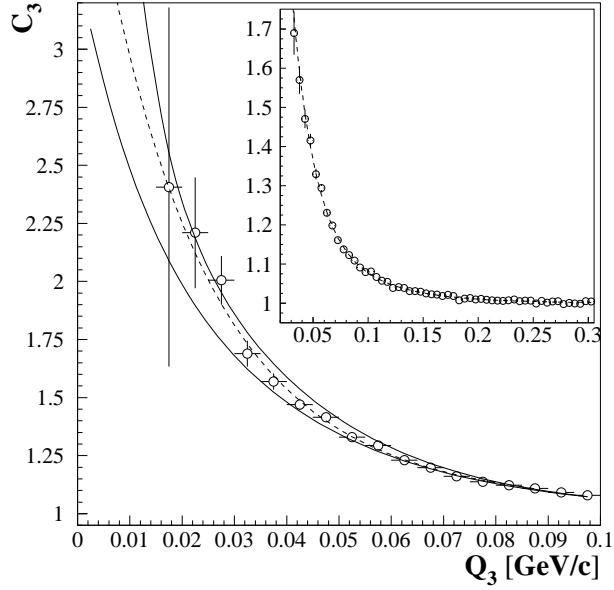}
}
\caption{The three-pion correlation function $C_3$ as a function of $Q_3$ 
with a fit to a double exponential form (dashed line--see text). 
The result is also compared to an estimate of $C_3$ with $W=1$
(upper solid line) and $W=0$ (lower solid line). The 
inset shows $C_3$ over a larger $Q_3$ range.}
\label{fig:figure2}
\end{figure}

Fig.~\ref{fig:figure2} shows the three-pion correlation as a function of
$Q_3$, after correction for resolution. A very strong $\pi^-\pi^-\pi^-$
correlation is observed, which
is robust under all tracking criteria.
The result shown is obtained for the 
same sample and the same cuts applied on the three pair combinations
contained in each triplet as used for the two-pion interference
analysis.
For the $C_3$ result, the Coulomb correction applied to a particular triplet is
the product of the Coulomb corrections used for the three pair
combinations contained in that triplet.
The resolution is taken into account 
using the same procedure
as in the two-pion
analysis.
The resolution has a tiny effect compared to the Coulomb correction and it
has been checked that the results are not affected by the order in which
these corrections are applied to the data.
 The dashed line is a fit to a
double exponential function \begin{equation}
C_3=1+\lambda_1\exp[-2Q_3R_1]+\lambda_2\exp[-2Q_3R_2]
\label{eq:c3fit}
\end{equation}
with fitted parameters $R_1=5.01\pm0.38$ fm, $\lambda_1=2.79\pm0.32$,
$R_2=1.72\pm0.12$ fm,
$\lambda_2=0.343\pm0.072$ and $\chi^2/d.o.f.=0.88$.
The three-pion correlation data cannot be well fitted by a Gaussian or a
single exponential as a function of $Q_3$.
Such non-Gaussian behaviour has been predicted by a final-state rescattering
model \cite{HUMAN}.

In Fig.~\ref{fig:figure2}, the three-pion correlation data are compared to
an estimate using Eq.~\ref{eq:c3exp} with 
$W=1$ (upper line) and $W=0$ (lower line).
This estimate is made using triplets from mixed events with the
$\lambda$ and
$R$ parameters extracted from the two-pion interferometry analysis.
Although the calculated contribution to $C_3$ from the 
genuine three-pion correlation is rather small 
and becomes insignificant for $Q_3>60-80$ MeV/c, the experimental
results clearly indicate a  
$W$ factor which lies between 0 and 1.

As proposed in Refs.~\cite{HEINZ,NA44}, 
a method to extract the experimental value of $W$ as a function of $Q_3$
is to invert Eqs.~\ref{eq:c3} and \ref{eq:cross} and rewrite
$|F_{ij}|$ in terms of $C_2$ using Eq.~\ref{eq:c2} to obtain 
\begin{equation}
W=\frac{ \{C_3(Q_3)-1\}-\{C_2(Q_{12})-1\}-\{C_2(Q_{23})-1\}-
\{C_2(Q_{31})-1\} }
   { 2\cdot\sqrt{ \{C_2(Q_{12})-1\}\{C_2(Q_{23})-1\}\{C_2(Q_{31})-1\} } }
\label{eq:w}
\end{equation}
In this method, the analysis must be performed in two steps.
First, the $\lambda$ and $R$ parameters are determined both for the
two-pion and three-pion correlations with a fit to the data of 
Eqs.~\ref{eq:c2exp} and \ref{eq:c3fit} as previously explained.
Then the data are analyzed again, and, for each triplet found,
characterized by the value $Q_3$, $W$ is determined 
using Eqs.~\ref{eq:c2exp}, \ref{eq:c3fit},  and \ref{eq:w} with the values
$Q_{12}$, $Q_{23}$, and $Q_{31}$ corresponding to the three pair combinations
contained in the triplet.
For each bin in $Q_3$ containing $N$ triplets, the statistical error on
the
mean value $\left<W\right>$ is $\sigma/\sqrt{N}$, where $\sigma^2$ is the
variance of the $W$ distribution in this particular bin.
The estimate of systematic errors is done by varying the different
analysis cuts in the two and three-pion interference studies.
It includes in particular the cuts used to identify the pions with the 
time of flight system.
The systematic error on the Coulomb correction due to the error
on the determination of the R parameter in the two-pion fit, as
well as a possible 10\% systematic error on the evaluation of the
$Q_{12}$ and $Q_3$ resolution are also taken into account.
The effects on $W$ of the statistical errors in the determination of $C_2$
and $C_3$ are treated as systematic errors by changing $C_2$ and $C_3$
respectively by $\pm\sigma_{C_2}$ and $\pm\sigma_{C_3}$.
All of these variations are then added in quadrature.

\begin{figure}[b]
\hspace{4.5cm}
\resizebox{0.5\textwidth}{!}{%
  \includegraphics{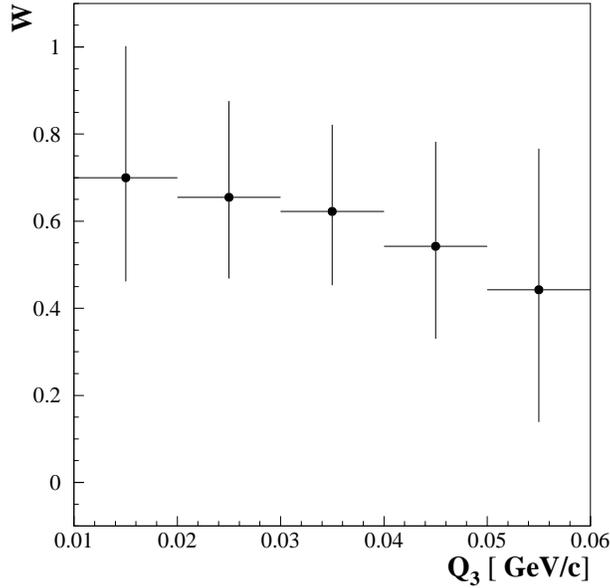}
}
\caption{The factor $W$ as a function of $Q_3$. The error bars include 
statistical and systematic errors. The statistical errors alone (not shown) 
are contained within the size of the symbols except for the first bin where 
it amounts to twice the size of 
the symbol. The horizontal bars indicate the bin width.}
\label{fig:figure3}
\end{figure}

Fig.~\ref{fig:figure3} shows the $W$ values obtained for five bins of
10 MeV/c in $Q_3$.
The error bars are the sum of statistical and systematic errors.
The statistical errors (not shown separately in Fig.~\ref{fig:figure3})
are nearly negligible.
The slight $Q_3$ dependence
observed is not significant in view of the errors.
The genuine three-body correlation is substantial with a weighted mean
$\left<W\right>=0.606\pm0.005\pm0.179$ for $Q_3<60$ MeV/c.
\footnote{
The weighted systematic error is obtained by calculating the weighted
average
over the five $Q_3$ bins separately for each kind of systematic error. 
These errors are then added in quadrature.
On the contrary, adding quadratically the systematic errors of the five 
$Q_3$
bins as done for
weighted statistical errors, or as done in ~\cite{NA44}, would give
$\pm0.097$ instead of $\pm0.179$ for the systematic uncertainty.}
This result should be compared to
the lower $\pi^+\pi^+\pi^+$
result of $\left<W\right>=0.20\pm0.02\pm0.19$ observed by the 
NA44 collaboration~\cite{NA44} in S+Pb
minimum bias collisions at 200 GeV per nucleon.

In conclusion, we have studied the $\pi^-\pi^-\pi^-$ interference of pions
produced in central Pb+Pb collisions at 158 GeV per nucleon.
Although its contribution is small, the genuine three-pion correlation,
the portion of the correlation not trivially due to the two-pion
interference, has been extracted and found to be substantial. 
The genuine three-pion correlation is greater than reported
for S+Pb minimum bias collisions~\cite{NA44},
but not as large as expected for a fully chaotic and symmetric source.


We would like to thank the CERN-SPS accelerator crew
for providing an excellent Pb beam. 
This work was supported jointly by the German BMBF and DFG, the U.S. 
DOE, the Swedish NFR, the Dutch Stichting FOM, the Swiss National Fund,
the Humboldt Foundation, the Stiftung f\"{u}r deutsch-polnische
Zusammenarbeit, the Department of Atomic Energy, the Department
of Science and Technology and the University Grants Commission of
the Government of India, the Indo-FRG Exchange Programme, the PPE
division of CERN, the INTAS under contract INTAS-97-0158, the
Polish KBN under the grant 2P03B16815, the Grant-in-Aid for Scientific
Research
(Specially Promoted Research \& International Scientific Research)
of the Ministry of Education, Science, Sports and Culture, JSPS
Research Fellowships for Young Scientists,
the University of Tsukuba Special Research Projects, and
ORISE.
ORNL is managed by UT-Battelle, LLC, for the U.S. Department of Energy
under contract DE-AC05-00OR22725.

%
%
\end{document}